\documentclass[aps,prd,showpacs,nofootinbib,twocolumn,10pt,superscriptaddress]{revtex4-1}
\usepackage{color,amsmath,amssymb,graphicx,latexsym,lineno}
\usepackage{threeparttable,multirow,txfonts,subfigure,booktabs}

\usepackage{accents,slashed,ulem}

\newlength{\dhatheight}
\newcommand{\doublehat}[1]{%
    \settoheight{\dhatheight}{\ensuremath{\hat{#1}}}%
    \addtolength{\dhatheight}{-0.15ex}%
    \hat{\vphantom{\rule{1pt}{\dhatheight}}%
    \smash{\hat{#1}}}}
\graphicspath{{figs/}}

\begin{document}

\title{Exploring Lorentz Invariance Violation from Ultra-high-energy Gamma Rays Observed by LHAASO}

\author{Zhen Cao}
\affiliation{Key Laboratory of Particle Astrophyics \& Experimental Physics Division \& Computing Center, Institute of High Energy Physics, Chinese Academy of Sciences, 100049 Beijing, China}
\affiliation{University of Chinese Academy of Sciences, 100049 Beijing, China}
\affiliation{TIANFU Cosmic Ray Research Center, Chengdu, Sichuan,  China}

\author{F. Aharonian}
\affiliation{Dublin Institute for Advanced Studies, 31 Fitzwilliam Place, 2 Dublin, Ireland }
\affiliation{Max-Planck-Institut for Nuclear Physics, P.O. Box 103980, 69029  Heidelberg, Germany}

\author{Q. An}
\affiliation{State Key Laboratory of Particle Detection and Electronics, China}
\affiliation{University of Science and Technology of China, 230026 Hefei, Anhui, China}

\author{Axikegu}
\affiliation{School of Physical Science and Technology \&  School of Information Science and Technology, Southwest Jiaotong University, 610031 Chengdu, Sichuan, China}

\author{L.X. Bai}
\affiliation{College of Physics, Sichuan University, 610065 Chengdu, Sichuan, China}

\author{Y.X. Bai}
\affiliation{Key Laboratory of Particle Astrophyics \& Experimental Physics Division \& Computing Center, Institute of High Energy Physics, Chinese Academy of Sciences, 100049 Beijing, China}
\affiliation{TIANFU Cosmic Ray Research Center, Chengdu, Sichuan,  China}

\author{Y.W. Bao}
\affiliation{School of Astronomy and Space Science, Nanjing University, 210023 Nanjing, Jiangsu, China}

\author{D. Bastieri}
\affiliation{Center for Astrophysics, Guangzhou University, 510006 Guangzhou, Guangdong, China}

\author{X.J. Bi$^*$}
\affiliation{Key Laboratory of Particle Astrophyics \& Experimental Physics Division \& Computing Center, Institute of High Energy Physics, Chinese Academy of Sciences, 100049 Beijing, China}
\affiliation{University of Chinese Academy of Sciences, 100049 Beijing, China}
\affiliation{TIANFU Cosmic Ray Research Center, Chengdu, Sichuan,  China}

\author{Y.J. Bi}
\affiliation{Key Laboratory of Particle Astrophyics \& Experimental Physics Division \& Computing Center, Institute of High Energy Physics, Chinese Academy of Sciences, 100049 Beijing, China}
\affiliation{TIANFU Cosmic Ray Research Center, Chengdu, Sichuan,  China}

% ...

\collaboration{LHAASO Collaboration}

\email[E-mail: ]{gaolq@ihep.ac.cn; chenes@ihep.ac.cn; bixj@ihep.ac.cn; yuanq@pmo.ac.cn; zhangyi@pmo.ac.cn; zhaosp@mail.sdu.edu.cn}

\date{\today}

\begin{abstract}
Recently the LHAASO Collaboration published the detection of 12 ultra-high-energy gamma-ray sources above 100 TeV, with the highest energy photon reaching 1.4 PeV. The first detection of PeV gamma rays from astrophysical sources may provide a very sensitive probe of the effect of the Lorentz invariance violation (LIV), which results in decay of high-energy gamma rays in the superluminal scenario and hence a sharp cutoff of the energy spectrum. Two highest energy sources are studied in this work. No signature of the existence of LIV is found in their energy spectra, and the lower limits on the LIV energy scale are derived. Our results show that the first-order LIV energy scale should be higher than about $10^5$ times the Planck scale $M_{pl}$ and that the second-order LIV scale is $>10^{-3}M_{pl}$.
Both limits improve by at least one order of magnitude the previous results.
\end{abstract}

\maketitle

\section{introduction}
  
The Lorentz invariance (LI) is one of the fundamental principles of the special relativity theory. However, many extensions of the standard 1model (SM) of particle physics, especially those trying to unify quantum mechanics and general relativity \cite{QG_S:1,QG_S:2,QG_S:3,QG_S:4,QG_S:5,QG_S:6,QG_S:7,QG_S:8}, suggest the Lorentz invariance violation (LIV) at the energy scale approaching the Planck scale $M_{pl}$. The LIV effect at low energies should be so tiny to be consistent with large amount of observations, but it may appear at very high energies which can be probed by observations of ultra-high-energy cosmic rays and gamma rays.

At low energies the LIV interaction can be expressed as an effective model by introducing LIV terms in the SM Lagrangian. These LIV terms will modify the particle dispersion relation, altering the standard on-shell condition of a particle energy-momentum relation in special relativity. As a result of the modified dispersion relation (MDR), the kinematics of particle propagation in the vacuum and particle interactions changes. Interesting phenomena which are forbidden in special relativity can occur with the MDR, such as the vacuum Cherenkov emission of charged particles, and the birefringence, decay or splitting of photons when propagating in the vacuum.

Astrophysical sources are ideal targets to search for the LIV effects because extremely high-energy processes can occur in these objects and the long distance to the Earth may result in an accumulation of the tiny effect. There have been many studies to explore the effects induced by LIV, such as the energy-dependent time delay from pulsars \cite{Pulsar_LIV_2017}, gamma-ray bursts (GRBs) \cite{GRB_fermi, Vasileiou:2013vra, ma_2018}, and flaring active galactic nuclei (AGN) \cite{AGN_Abdalla_2019,AGN_LIV_Abramowski_2011}, the vacuum Cherenkov emission \cite{Galaverni_2008_Cherenkov1, Gagnon_2004_Cherenkov2}, the vacuum birefringence \cite{vac_birefringence_10.1093/mnras/stw2007, vac_birefringence_10.1093/mnras/stz594}, and the decay or splitting of photons \cite{Satunin:2019gsl, PhysRevD.95.063001, CANGAROO:Crab, Albert:2019nnn}.

The MDR of a photon can be written as \cite{MDR1_Coleman_1999, MDR2_Amelino_Camelia_1998, MDR3_AmelinoCamelia:2001dy, MDR4_Ahluwalia_1999}
\begin{equation}
E_\gamma^2 - p_\gamma^2 = \pm  |\alpha_n| p_\gamma^{n+2}  , 
\label{mdr}
\end{equation}
where $E_\gamma$ and $p_\gamma$ are the energy and momentum of a photon, $\pm$ corresponds to superluminal ($+$) and subluminal ($-$) cases. For $n \textgreater 0$, $\alpha_n$ is the $n$th order LIV parameter which is related to the LIV energy scale, i.e., $E_{\rm{LIV}}^{(n)} = \alpha_n^{-1/n}$.

In this work, we study the superluminal LIV effect in photons using the observations of unprecedentedly high energy $\gamma$ rays by the Large High Altitude Air Shower Observatory (LHAASO) \cite{LHAASO:whitepaper}. In the superluminal LIV case, photons can decay into a pair of electron and positron, $\gamma\to e^-e^+$, as long as the threshold condition is satisfied. This process occurs rapidly and leads to a sharp cutoff in the $\gamma$-ray spectrum \cite{PhysRevD.95.063001, Mart_nez_Huerta_2016, Martinez-Huerta:2017ntv}. The superluminal LIV would also lead to a $\gamma$-ray photon splitting into multiple photons, $\gamma\to N\gamma$. The dominant process of the photon splitting is $\gamma\to 3\gamma$ \cite{Astapov:2019xmt, article111}. Although there is no threshold energy for the photon splitting process, it also results in a hard cutoff in the $\gamma$-ray spectrum because the decay width depends heavily on the photon energy \cite{Astapov:2019xmt, Satunin:2019gsl,article111}.

Recently, a list of 12 $\gamma$-ray sources detected with more than $7\sigma$ at  energies above 100 TeV were reported by the LHAASO collaboration \cite{LHAASO_nature}. The highest energy $\gamma$-like event, from source LHAASO J2032+4102, is about 1.4 PeV. The second highest energy $\gamma$-like event, from source LHAASO J0534+2202 (Crab Nebula), is about 0.88 PeV. 
The measurements do not show clear cutoff at the highest energy end in their spectra, and thus stringent constraints on the LIV energy scale can be derived using the data \cite{2021arXiv210507927C}. 
In this work, we study the superluminal LIV effect using the LHAASO data, with a rigorous statistical approach and a careful assessment of the systematics. Since the highest energy photons may provide the most stringent constraints, we only use the two sources LHAASO J2032+4102 and J0534+2202 in this study.

\section{The LHAASO experiment and the data}	

\subsection{LHAASO}
LHAASO is a new generation gamma-ray and cosmic-ray observatory, which is under construction at an altitude of 4410 m with location $29^\circ 21' 31''$ N, $100^\circ 08' 15''$ E in Daocheng, Sichuan province, China \cite{LHAASO:whitepaper}. LHAASO consists of three detector arrays, the Kilometer Square Array (KM2A), the Water Cherenkov Detector Array (WCDA), and the Wide Field-of-view Cherenkov Telescope Array (WFCTA). A large fraction of the LHAASO detectors started the operation since 2019, and the whole detector construction will be finished in 2021 \cite{LHAASO:KM2A_perf}.

KM2A is composed of 5195 electromagnetic detectors (EDs) and 1188 muon detectors (MDs), which cover an area of 1.3 $\rm{km}^2$. EDs (MDs) are distributed with a spacing of 15 m (30 m). In the outskirt ring region, additional EDs with interval of 30 m are placed to discriminate showers with cores located inside and outside the central region. KM2A has a wide field-of-view (FOV) of $\sim2$ sr and observes 60\% of the sky with one day exposure. It provides an unprecedented sensitivity to survey the $\gamma$-ray sky for energies above 20 TeV.

\subsection{Data}

Data of the half array of LHAASO-KM2A from December 26, 2019, to November 30, 2020, corresponding to a live time of about 301.7 days, is used in this work. The detection efficiency of a typical ED (MD) is about 98\% (95\%) and the time resolution of ED (MD) is about 2 ns (10 ns). Hits of EDs are used to reconstruct the direction, core, and energy of primary particles. 
MDs are used to discriminate  $\gamma$-ray induced showers from showers generated by cosmic rays. We adopted a 400 ns time window and a 100 m (radius) spatial window to select the shower hits. The core location and direction of the shower can be obtained through a fitting to the shower front with a modified Nishimura-Kamata-Greisen (NKG) function \cite{NKG}. A robust estimator $\rho_{50}$, defined as the particle density that best-fits the modified NKG function at a perpendicular distance of 50 m from the shower axis, is adopted to reconstruct the energy. The core resolution (68\% containment) is about 4-9 m (2-4 m) at 20 (100) TeV, the angular resolution is $0^\circ\!.5 - 0^\circ\!.8$ ($0^\circ\!.24 -  0^\circ\!.3$) at 20 (100) TeV depending on the zenith angle, and the energy resolution is about 24\% (13\%) at 20 (100) TeV for showers with zenith angles smaller than $20^\circ$ \cite{LHAASO:KM2A_perf}. For the PeV energy photons which are most relevant in this study, the energy resolution can reach $\sim8.5\%$ for zenith angles smaller than $20^\circ$ \cite{LHAASO:KM2A_perf}.

Showers induced by cosmic rays have more muons than those induced by photons. So we can reject the cosmic ray background through the ratio $N_{\mu}/N_e$, where $N_{\mu}$ is the number of muons and $N_e$ is the number of electromagnetic particles. If we keep a 90\% efficiencies for primary $\gamma$ rays, the cosmic ray background can be rejected by 99\%, 99.99\%, and 99.997\% at 20, 100, and 1000 TeV energies, respectively \cite{LHAASO:KM2A_perf}.
Additional selections require zenith angles smaller than $50^\circ$, shower ages within 0.6 to 2.4, and both the numbers of fired EDs and secondary particles used for reconstruction larger than 10.

The sky map in the celestial coordinate (right ascension and declination) is divided into a grid of $0^\circ\!.1 \times 0^\circ\!.1$ pixels, filled with events detected by KM2A. The background in each pixel can be estimated through the ``direct integration method'' \cite{Fleysher_2004}. This method estimates the background of one pixel by using events in the same pixel in the local coordinate but at different time. In this work, events accumulated in eight hours are integrated to estimate the detector acceptance for each pixel in this time interval \cite{Fleysher_2004}.
This method can reduce the influence of instrumental and environmental variations. 

\section{Method}

Gamma-like events from directions of Crab Nebula and LHAASO J2032+4102 are used in this analysis. Crab Nebula is a pointlike source with the KM2A resolution, and LHAASO J2032+4102 is found to be extended with a Gaussian width of $0^\circ\!.3$.  The analysis of the energy spectra of the sources is similar with that of Ref.~\cite{LHAASO:KM2A_perf}.
Slight optimizations are employed to better estimate the LIV cutoff value. 
First, we re-bin energies with a width of $\Delta {\rm log}_{10}E=0.1$ ranging from 10 TeV to 1.58 PeV. This bin width is smaller than the LHAASO energy spread at these energies. We find no significant difference for LIV limit when using a finer bin width, e.g., $\Delta{\rm log}_{10}E=0.05$. Second, we improve the background estimation. For the highest energy bins, the statistics is too low to estimate the detector acceptance precisely
with eight hours' data. 
We stack the events in one sidereal day within a larger off-region
to estimate the background.

Two spectral forms are assumed for both sources, the log-parabolic form and the power-law form. Both $\gamma \to e^+ e^-$ and $\gamma \to 3\gamma$ processes predict that the energy spectrum of a source has a quasi-hard cutoff. Therefore the expected spectrum of these sources, when there is LIV, is
\begin{equation}
f(E)=\phi_0 \left(\frac{E}{E_0}\right)^{-\alpha-\beta {\rm ln}(E/E_0)} H(E-E_{\rm{cut}})\ \, ,
\label{equ:sed1}
\end{equation}
where $\phi_0$, $\alpha$, and $\beta$ are flux normalization and spectral indices, $E_0 = 20 $ TeV is a reference energy, $H(E-E_{\rm{cut}})$ is the Heaviside step function, and $E_{\rm{cut}}$ is the cutoff energy. 
The above formula is for the log-parabolic spectrum, and we set $\beta\equiv0$ for the power-law spectrum.

We use the forward folding procedure to get the energy spectra. The spectral parameters are obtained based on the maximum likelihood fitting algorithm. The likelihood function is defined as
\begin{equation}
\mathcal{L}(\phi_0, \alpha, \beta, E_{\rm{cut}}) = \prod_{i = 1}^{n} {\rm Poisson}(N^{i}_{\rm obs}, N_{\rm sig}^i(\phi_0, \alpha, \beta, E_{\rm cut}) + N_{\rm bkg}^i), 
\end{equation}
where $i$ denotes the $i$-th energy bin, $N^{i}_{\rm obs}$ is the number of observed events from the source, $N_{\rm bkg}^i$ is the estimated background, and $N_{\rm sig}^i$ is the expected signal calculated by convolving the spectrum with the KM2A energy resolution. For each $E_{\rm cut}$, we can get the best-fit spectral parameters $\phi_0, \alpha, \beta$, and the corresponding likelihood value.

The signiﬁcance of the existence of such a hard cutoff was estimated using a test statistic (TS) variable, which is the logarithm of the likelihood ratio of the fit with a cutoff $E_{\rm cut}$ and the fit with $E_{\rm cut} \to \infty$, 
\begin{align}
{\rm TS}(E_{\rm cut}) = -\sum_{\rm bin} 2\ln\left(\frac{\mathcal{L}_1( \doublehat{\phi}_0, \doublehat{\alpha}, \doublehat{\beta} , E_{\rm cut})}{\mathcal{L}_0(\hat{\phi}_0, \hat{\alpha}, \hat{\beta}, E_{\rm cut} \to \infty)}\right).
\end{align}
The null hypothesis (without the LIV effect) corresponds to $E_{\rm cut}\to \infty$, and the alternative hypothesis (with the LIV effect) is the case with a finite $E_{\rm cut}$.

The spectral fit does not favor the existence of a cutoff for both sources. Therefore a lower limit on $E_{\rm cut}$ can be set, below which photons should not decay. Since the statistics in the highest energy bins is rather poor, the TS value does not follow a $\chi^2$ distribution \cite{Algeri2020_wilk}. In this case, the Wilks’ theorem \cite{10.1214/aoms/1177732360} is not appropriate to estimate the confidence level (CL) of $E_{\rm cut}$. Hence, we adopt the $CLs$ method \cite{Read_2002} to derive the 95\% CL limit of $E_{\rm cut}$.

\begin{figure}
	\centering
	\includegraphics[width=8.6cm]{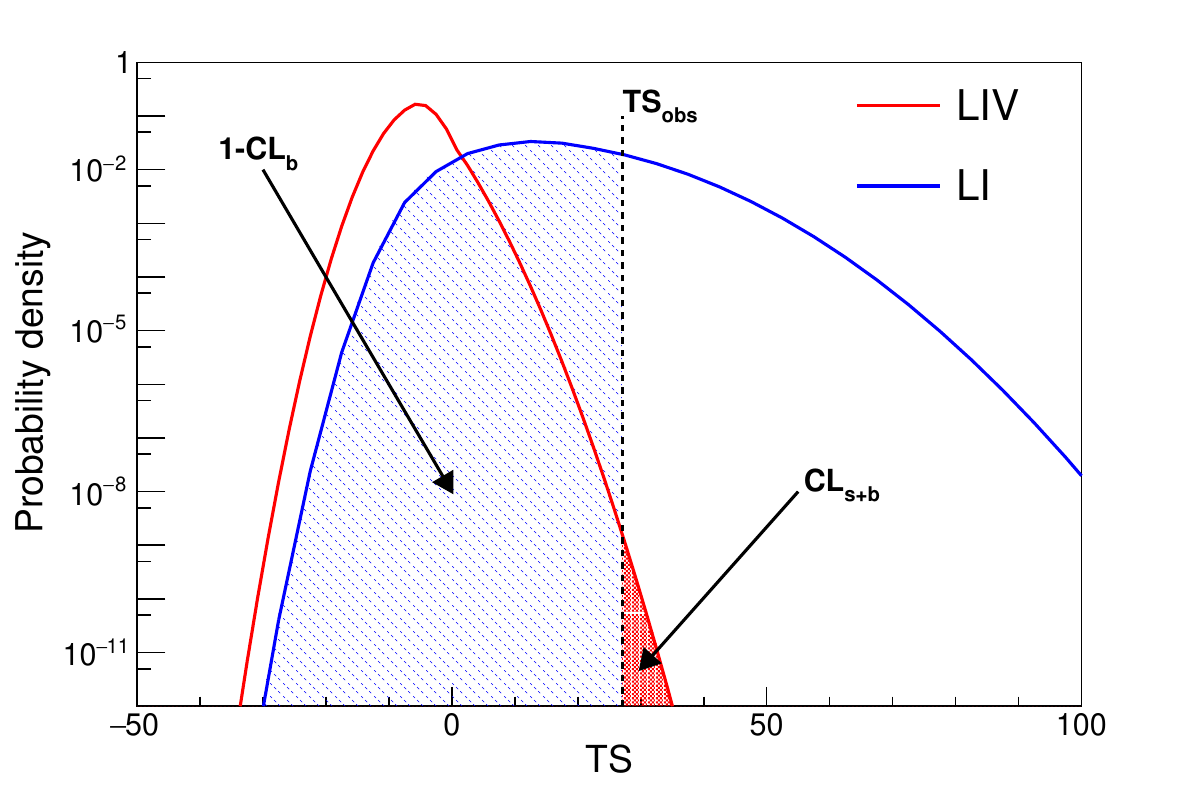}
	\caption{TS distributions of MC data with and without spectral cutoff, for Crab Nebula and $E_{\rm cut} = 250$ TeV. The red line is for the LIV case, and the blue line is for the LI case.}
	\label{fig:TS_dis}
\end{figure}

\begin{figure}
	\centering
	\includegraphics[width=8.6cm]{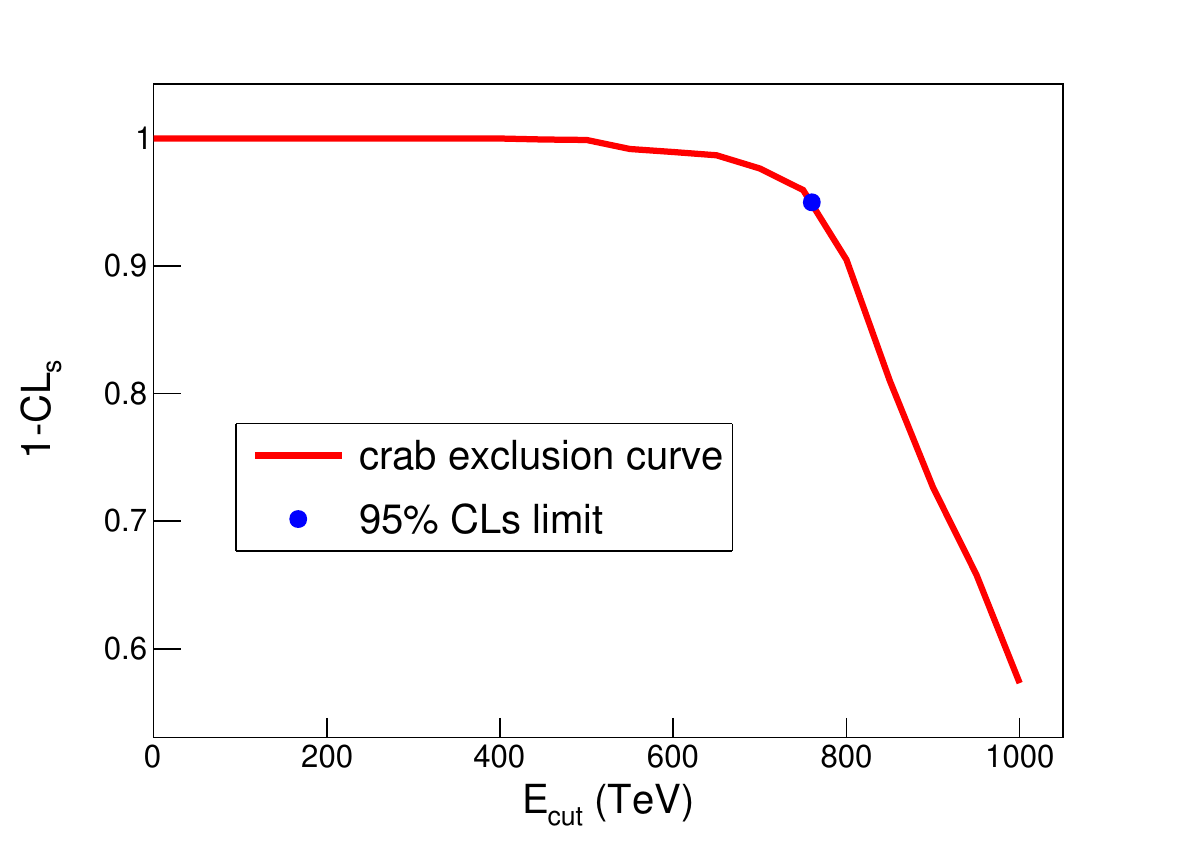}
	\caption{The probability $1-CL_s$ as a function of $E_{\rm cut}$ for  Crab Nebula. The blue dot marks the $E_{\rm cut}$ value which is excluded at the 95\% CL with the $CL_s$ method.}     \label{fig:cls_ecut}
\end{figure}

The probability distribution of the TS values for the null hypothesis ($E_{\rm cut} \to \infty$) and the signal plus background hypothesis are obtained using Monte Carlo (MC) simulations. The background $N_{\rm bkg}^i$ is obtained from the experimental data. For a given $E_{\rm cut}$, we calculate the corresponding TS values for both MC data sets with and without the LIV. As an example, Fig. \ref{fig:TS_dis} shows the TS distributions for Crab Nebula for $E_{\rm cut}=250$ TeV. The red (blue) line is the TS distribution derived from the MC data with (without) the LIV effect. The TS value derived from the observational data ${\rm TS}_{\rm obs}(E_{\rm cut}=250$ $\rm{TeV})$ is $27.2$. The red shaded region indicates the probability for ${\rm TS} \geq {\rm TS}_{\rm obs}$ under the 
hypothesis of $E_{\rm cut} = 250$ TeV, defined as $CL_{s+b}$. The probability for ${\rm TS} \leq {\rm TS}_{\rm obs}$ under the $E_{\rm cut} \to \infty$ hypothesis is defined as $1-CL_{b}$, as indicated by the blue shaded region.  The definition of $CL_s$ is $CL_s = CL_{s+b}/CL_{b}$. If $CL_s<0.05$, the LIV scenario with a certain $E_{\rm cut}$ value is excluded at the 95\% CL. Fig. \ref{fig:cls_ecut} shows the $CL_s$ as a function of $E_{\rm cut}$ for Crab Nebula. The blue point is derived as the 95\% CL lower limit of $E_{\rm cut}$.

\section{results}

\begin{figure}
	\centering
	\includegraphics[width=8.6cm]{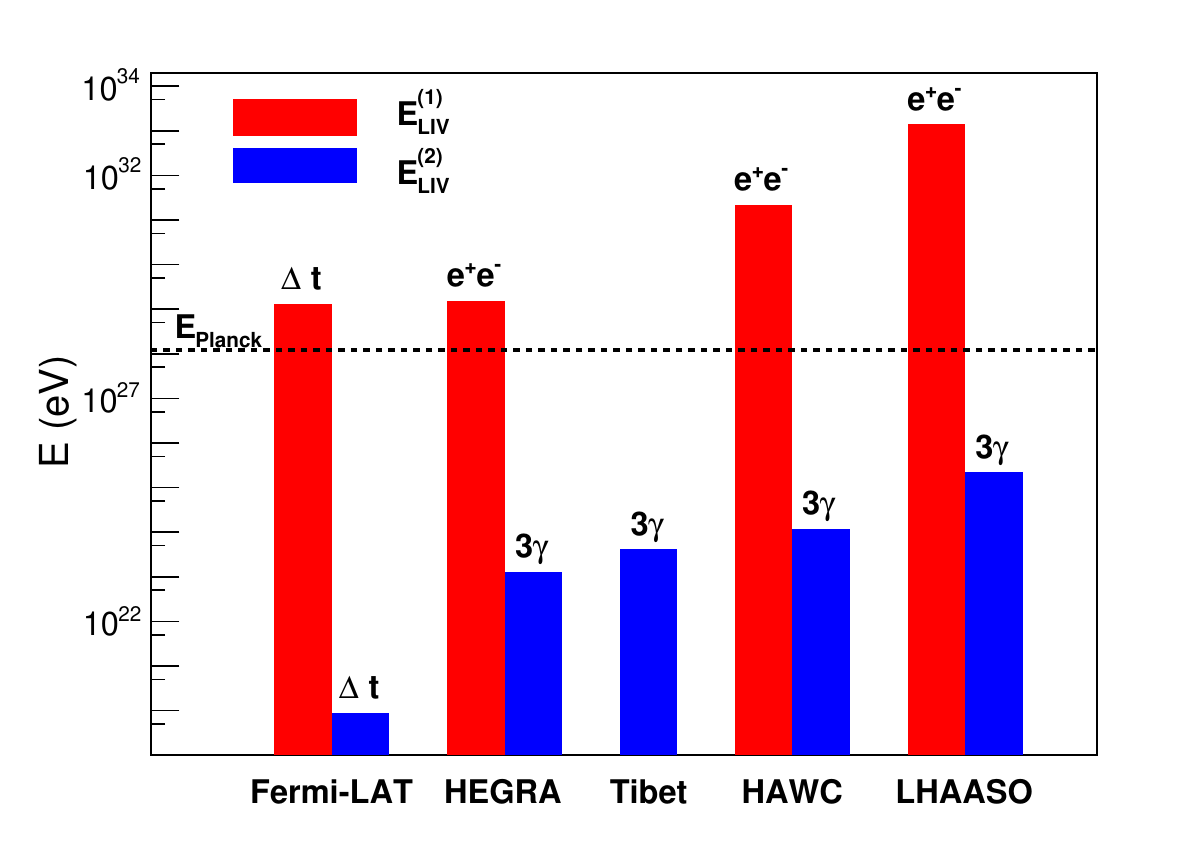}
	\caption{Comparison of the constraints on the $E^{(1)}_{\rm LIV}$ and $E^{(2)}_{\rm LIV}$ derived from LHAASO and other experiments \cite{Astapov_2019,PhysRevD.95.063001,Satunin:2019gsl,Albert:2019nnn,Vasileiou:2013vra}. We show constraints due to the photon decay ($e^+e^-$) and the photon splitting ($3\gamma$) processes for all experiments except for Fermi-LAT which adopted the time delay method ($\Delta t$).}
	\label{fig:comp}
\end{figure}

Adopting the $CL_s$ method introduced above we derive the 95\% CL lower limits on the LIV induced cutoff energy $E_{\rm cut}$ in spectra based on the LHAASO-KM2A data concerning the two highest energy sources  LHAASO J0534+2202 and LHAASO J2032+4102.
Table \ref{tab:Ec_res} lists the 95\% CL limits on $E_{\rm cut}$ and the inferred LIV energy scales. The 95\% CL lower limits for the cutoff energy are 750 TeV and 1140 TeV for LHAASO J0534+2202 and LHAASO J2032+4102, respectively. The LIV energy scale limits are derived from $E_{\rm cut}$ following the formulas in \cite{PhysRevD.95.063001,Albert:2019nnn} for the two processes. The combined limit from the two sources for the process $\gamma \to e^+e^-$ is also derived and the result is nearly the same as 1140 TeV. For the process $\gamma \to 3 \gamma$, the LIV energy scale depends on the distances of the sources weakly and no combined result is given. 
In the direction of LHAASO J2032+4102, there are more than one potential counterparts \cite{LHAASO_nature}.
We adopt the smallest distance of the potential counterparts for a conservative estimate. For other distance the limit changes slightly. The combined limit on the first-order LIV energy scale is about $1.42 \times 10^{33}~\rm{eV}$, which is five orders of magnitude higher than the Planck scale ($M_{pl}\approx1.22\times10^{28}$~eV). The second-order LIV energy scale reaches $10^{-3}$ times of the Planck scale, as derived from the $\gamma \to 3 \gamma$ process. The comparison with the results obtained from other experiments is shown in Fig. \ref{fig:comp}. We show the limits on the decay of photons and the photon splitting from the HEGRA \cite{Astapov_2019,PhysRevD.95.063001}, Tibet \cite{Satunin:2019gsl}, and HAWC \cite{Albert:2019nnn} observations.  The limit from analyzing Fermi-LAT observations of energy-dependent time delays of GRB photons is also shown \cite{Vasileiou:2013vra}. Our results improve by more than one order of magnitude the previous results, and give by far the most stringent constraints on the energy scales of the superluminal LIV.

\begin{table*}[htbp]
	\caption{Columns are sources we studied, distances, the highest energies of photons recorded by LHAASO-KM2A, the 95\% CL lower limits on $E_{\rm cut}$, lower limits on the first and second order LIV scales derived from $E_{\rm cut}$ for the process $\gamma \to e^+ e^-$ , and lower limits on the second order LIV scale from process $\gamma \to 3 \gamma$. The systematic errors on the derived values are also shown. } 
	\centering 
	\renewcommand\tabcolsep{10.0pt}

	\begin{tabular}{c ccccccccc} 
		\hline\hline 
		Source & $L$ & $E_{\rm max}$ & $E_{\rm cut}^{95\%}$ & $E_{\rm LIV}^{(1)}$ & $E_{\rm LIV}^{(2)}$ & $E_{\rm LIV}^{(2)}$ $(3\gamma)$\\
		& (kpc) & (PeV) & (PeV) &  (eV) & (eV) & (eV) \\ 
		& & & & $\times 10^{32}$ & $\times 10^{23}$ &  $\times 10^{25}$ \\ [0.5ex]
		\hline      
		J0534+2202 & 2.0 & 0.88 & $0.75^{+0.043}_{-0.043}$ & $4.04^{+0.73}_{-0.65}$ & $5.5^{+0.65}_{-0.61}$ & $1.04^{+0.12}_{-0.11}$ \\ 
		J2032+4102 & 1.4 & 1.42 & $1.14^{+0.06}_{-0.06}$ & $14.2^{+2.32}_{-2.10}$ & $12.7^{+1.36}_{-1.29}$ & $2.21^{+0.22}_{-0.21}$ \\ [0.5ex]
		\hline
		\hline 
	\end{tabular}
	\label{tab:Ec_res}
\end{table*}

\section{Systematic uncertainties}

There are several systematic uncertainties that affect the LIV energy scale constraints. 
In one year's operation, some percent of detector units was occasionally switched to the debug mode, and thus the layout of the array varied slightly with time. Furthermore, uncertainties in the modeling of the atmosphere may affect the simulation results. These effects lead to the flux and spectral index of the energy spectrum varying by about 7\% and 0.02, respectively. The uncertainties on the spectral parameters would lead to a 1.5\% effect on the $E_{\rm cut}^{95\%}$ value.

The assumed spectral model also leads to a systematic uncertainty. We compare results by adopting different spectral models, the log-parabolic,  power-law and broken power-law models, and find a $\sim 5$\% difference. The combined systematic uncertainty is estimated to be about $5.2\%$.
These uncertainties lead to an error on deriving $E_{\rm cut}^{95\%}$ and the corresponding LIV scales given in Table \ref{tab:Ec_res}.

\section{Summary}
Twelve sources above 100 TeV were detected with high significance by LHAASO-KM2A. Among them, LHAASO J0534+2202 and LHAASO J2032+4102 are the two sources with the highest energy $\gamma$-like events up to PeV energies. The ultra-high-energy $\gamma$ events are used to constrain the LIV effect, which is predicted to give hard cutoff to the energy spectra of $\gamma$-ray sources due to the MDR-induced photon decay or splitting.
To get a precise 95\% CL lower limit on $E_{\rm cut}$, pseudo-experiments by MC simulations are carried out and the $CL_s$ method is adopted.  The first-order LIV energy-scale is constrained to be higher than $10^5 M_{pl}$, and the second-order LIV energy-scale should exceed $10^{-3} M_{pl}$. These results are the strongest constraints on the superluminal LIV parameters among experimental results with similar technique.

\acknowledgements

This work is supported in China by the National Key R$\&$D program of China under the grants 2018YFA0404202, 2018YFA0404201, 2018YFA0404203, 2018YFA0404204, the National Natural Science Foundation of China under the grants 11635011, 11761141001, 11765019, 11775233, U1738205, U1931111, U1931201, U2031201, Chinese Academy of Sciences, and the Program for Innovative Talents and Entrepreneur in Jiangsu, and in Thailand by RTA6280002 from Thailand Science Research and Innovation. The authors would like to thank all staff members who work at the LHAASO site which is 4400 meters above sea level year-around to maintain the detector and keep the electricity power supply and other components of the experiment operating smoothly. We are grateful to Chengdu Management Committee of Tianfu New Area for the constant ﬁnancial supports to research with LHAASO data.

\bibliographystyle{apsrev}
\bibliography{refs}

\end{document}